\documentclass[3p,times]{elsarticle}
\usepackage{ecrc}

\volume{00}

\firstpage{1}

\journalname{Annals of Physics}

\runauth{S. Longhi}

\usepackage{amssymb}
\usepackage[figuresright]{rotating}

\begin{document}

\begin{frontmatter}



\title{Non-Hermitian time-dependent perturbation theory: asymmetric transitions and transitionless interactions}


\author{Stefano Longhi and Giuseppe Della Valle}
\address{Dipartimento di Fisica, Politecnico di Milano and Istituto di Fotonica e Nanotecnologie del Consiglio Nazionale delle Ricerche, Piazza L. da Vinci 32, I-20133 Milano, Italy\\ Tel/Fax: 0039 022399 6156/6126, email: longhi@fisi.polimi.it}

\begin{abstract}
The ordinary time-dependent perturbation theory of quantum mechanics, that describes the interaction of a stationary system with a time-dependent perturbation, predicts that the transition probabilities induced by the perturbation are symmetric with respect to the initial an final states. Here we extend time-dependent perturbation theory into the non-Hermitian realm and consider the transitions in a stationary Hermitian system, described by a self-adjoint Hamiltonian $\hat{H}_0$, induced by a time-dependent non-Hermitian interaction $f(t) \hat{H}_1$. In the weak interaction (perturbative) limit, the transition probabilities  generally turn out to be {\it asymmetric} for exchange of initial and final states. In particular, for a temporal shape $f(t)$ of the perturbation with one-sided Fourier spectrum, i.e. with only positive (or negative) frequency components, transitions are fully unidirectional, a result that holds even in the strong interaction regime. Interestingly, we show that non-Hermitian perturbations can be tailored to be transitionless, i.e. the perturbation leaves the system unchanged as if the interaction had not occurred at all, regardless the form of $\hat{H}_0$ and $\hat{H}_1$.  As an application of our results, we provide important physical insights into the asymmetric (chiral) behavior of dynamical encircling of an exceptional point in two- and three-level non-Hermitian systems. 
\end{abstract}

\begin{keyword}
non-Hermitian dynamics; time-dependent perturbation theory; exceptional points
\end{keyword}

\end{frontmatter}

\section{Introduction}
Predicting and controlling the temporal evolution of a quantum system under the effect of a time-dependent 
perturbation is of central importance to a wide variety of problems in quantum physics, such as in quantum scattering, quantum control and quantum engineering, laser-driven atomic and molecular physics, and quantum information processing.
For the standard situation, i.e. when the Hamiltonian
is Hermitian, there exist well developed mathematical tools, such as time-dependent perturbation theory, Dyson series, adiabatic theory for slowly-changing parameters,  Floquet theory for periodic perturbations, etc. \cite{r1,r2,r3}. One of the simplest cases, which is treated at a simple level in any quantum mechanical textbook, is that of a weak perturbation that describes an interaction with finite duration. The effect of the 
interaction is to induce transitions among the different eigenstates $|n \rangle$, $m \rangle$ of the unperturbed (stationary) system, which are described by the transition probabilities $W_{n,m}$ and expressed by the Fermi golden rule \cite{r1,r3}. For a weak perturbation, a very general result is that the transition probabilities turn out to be symmetric, i.e. $W_{n,m}=W_{m,n}$.  In many physical problems, however, one deals with systems described by a non-Hermitian Hamiltonian. Non-Hermitian Hamiltonians are widely used as effective models to describe open quantum and classical systems \cite{r4,r5,r6}, or are introduced to provide complex extensions of the ordinary quantum mechanics such as in the $\mathcal{PT}$-symmetric quantum mechanics \cite{r7,r8,r8bis,r8tris}. The increasing interest devoted to non-Hermitian dynamics has motivated the extension of the arsenal of perturbation mathematical tools into the non-Hermitian realm \cite{r9,r10,r10bis,r11,r11bis,r12,r13,r13bis,r13tris,r14,r15,r16,r17,r18,r19,r19bis,r20,r21,r21bis,r22,r23,r23bis,r23tris,r23quatris,r23quintris,r24,r25,r26,r27,r28,r29,r30}. 
Several results have been found concerning extensions and breakdown of the adiabatic theorem \cite{ r10bis,r11bis,r12,r23quatris,r26,r27}, Berry phase \cite{ r10,r11,r13,r13bis,r16,r19bis,r20,r23bis,r23tris} and shortcuts to adiabaticity \cite{r22,r24,r25,r29,r30}. As compared to Hermitian Hamiltonians, non-Hermitian ones can show unusual spectral behavior, such as the appearance of exceptional points (EPs) corresponding to the coalescence of two (or more) eigenvalues and of corresponding eigenfunctions \cite{r31,r32,r33}.  A particularly intriguing behavior is found when encircling an EP. While ultraslow (quasi-static) encircling results in adiabatic evolution of the system and final flip of the states \cite{r15,r34}, non-adiabatic transitions lead to a chiral behavior, i.e. different final states are observed when encircling an
EP in a clockwise or a counter-clockwise direction \cite{r21,r21bis,r23,r28,r35bis,r35}. \par
In this work we devise another unusual behavior of non-Hermitian dynamics by considering  the transitions in a stationary Hermitian system, described by a self-adjoint Hamiltonian $\hat{H}_0$, induced by a time-dependent {\it non-Hermitian} interaction $f(t) \hat{H}_1$. In the weak interaction (perturbative) limit, it is shown rather generally that the transition probabilities $W_{n,m}$ and $W_{m,n}$ between stationary states $|n \rangle$ and $| m \rangle$ of $\hat{H}_0$ turn out to be {\it asymmetric}, i.e. $W_{n,m} \neq W_{m,n}$. In particular, for a temporal shape $f(t)$ of the perturbation with one-sided Fourier spectrum, i.e. with only positive (or negative) frequency components, transitions are fully unidirectional, i.e. $W_{n,m}=0$ while $W_{m,n} \neq 0$ (or viceversa), a result that holds {\em even in the strong interaction regime}. Interestingly, non-Hermitian perturbations can be tailored to become transitionless, i.e. the perturbation leaves the system unchanged as if the interaction had not occurred at all. As an application of the above results, we provide important physical insights into asymmetric transitions in the dynamical encircling of an EP of second and third order, and relate the onset of chiral behavior to the asymmetric transition probabilities.

\section{Transitions induced by non-Hermitian time-dependent perturbations}
Let us consider a stationary Hermitian system, described by a time-independent self-adjoint Hamitlonian $\hat{H}_0$, which interacts with its environment in such a way that the interaction is described by a non-Hermitian time dependent Hamiltonian $f(t) \hat{H}_1$, where $f(t)$ is a scalar and generally complex function of time $t$ and $\hat{H}_1$ is the time-independent perturbation operator. The time-dependent Hamiltonian of the system that describes non-Hermitian interaction thus reads
\begin{equation}
\hat{H}(t)= \hat{H}_0 +f(t) \hat{H}_1.
\end{equation}
For the sake of definiteness, we assume that $\hat{H}_0$ has a pure point spectrum comprising a finite number $N$ of energies $E_n= \hbar \omega_n$ with corresponding orthonormal eigenstates $|n \rangle$ ($n=1,2,...,N$). However, the results discussed in the present paper can be extended to the case where $\hat{H}_0$ shows an absolutely continuous energy spectrum as well as in the $N \rightarrow \infty$ limit. Energies and corresponding eigenstates are ordered such that $E_1 \leq E_2 \leq E_3 \leq ... \leq E_N$. Following a standard procedure of time-dependent perturbation theory \cite{r1,r2}, we expand the state of the system $ | \psi(t) \rangle $ in series of the eigenstates $|n \rangle$ of $\hat{H}_0,$ i.e. we set $|\psi(t) \rangle= \sum_l c_l(t) \exp(-i E_l t) |l \rangle$. From the Schr\"odinger equation (with $\hbar=1$)  $i \partial_t |\psi(t) \rangle = \hat{H}(t) | \psi(t) \rangle$, the following evolution equations for the amplitude probabilities $c_l(t)$ are readily found
\begin{equation}
i \frac{dc_l}{dt}= f(t) \sum_{s} (H_1)_{l,s}c_s \exp [i (\omega_l-\omega_s)t ]
\end{equation}
where $(H_1)_{l,s}= \langle l | \hat{H}_1 s \rangle$ are the matrix elements of the perturbation operator $\hat{H}_1$. In the Hermitian limit, $f(t)$ real and $\hat{H}_1^{\dag}=\hat{H}_1$, norm conservation implies $\sum_n |c_n(t)|^2=1$, however for a non-Hermitian interaction the norm is generally not conserved and thus the amplitude probabilities can become even larger than one. 
We assume that the interaction vanishes as $t \rightarrow \pm \infty$, namely we assume that $f(t) \rightarrow 0$ as $t \rightarrow \pm \infty$ sufficiently fast so as one can define the frequency spectrum 
\begin{equation}
F(\omega)=\int_{-\infty}^{\infty} f(t) \exp(i \omega t)
\end{equation}
 of the perturbation. Equation (2) can be formally integrated yielding
\begin{equation}
c_l(t)=c_l(-\infty)+\sum_s (H_1)_{l,s} \int_{-\infty}^t d\xi f(\xi) c_s(\xi) \exp[i(\omega_l-\omega_s)\xi]
\end{equation}
where $c_l(-\infty)$ define the initial state of the system at $t \rightarrow -\infty$. 
Equation (4) is especially useful in the weak interaction limit $f(t) \rightarrow 0$, where the solution $c_n(t)$ can be obtained by an iterative method starting with the unperturbed values $c_{l}(-\infty)$ as a first trial under the sign of the integral on the right hand side \cite{r1,r2}. Let us assume, for example, that before the interaction the system is prepared in the stationary state $|n \rangle$, i.e. $c_{l}(-\infty)=\delta_{l,n}$. The effect of the interaction is to induce a transition from state $| n \rangle$ into the other stationary states of $\hat{H}_0$. The transition probability $W_{n,m}$ from state $| n \rangle$ to state $|m \rangle$ is given by $W_{n,m}=|c_{m}( \infty)|^2$.  For a weak perturbation $f(t) \rightarrow 0$, a simple expression of $W_{n,m}$ can be obtained by first-order perturbation theory, which is simply obtained by letting $c_s(t)=\delta_{s,n}$ on the right hand side of Eq.(4). This yields
\begin{equation}
W_{n,m} \simeq |(H_1)_{m,n}|^2 \left| \int_{-\infty}^{\infty} dt f(t) \exp [i (\omega_m-\omega_n) t ] \right|^2=|(H_1)_{m,n}|^2  |F(\omega_m-\omega_n)|^2. 
\end{equation} 
The standard time-dependent perturbation theory is obtained by letting $f(t)$ real and $\hat{H}_1$ self-adjoint, i.e. $(H_1)_{n,m}=(H_1)_{m,n}^*$. In this case, from Eq.(5) and from the definition of the spectrum $F(\omega)$ of perturbation [Eq.(3)] one has $W_{n,m}=W_{m,n}$, i.e. the transition probability is symmetric for exchange of initial and final states. This is a well-known result in ordinary quantum mechanics and related to the Fermi golden rule result \cite{r1,r2,r3}. However, when the interaction is described by a non-Hermitian Hamiltonian, the transition probabilities clearly become rather generally asymmetric, i.e. $W_{n,m} \neq W_{m,n}$. For example, if $f(t)$ is real but $\hat{H}_1$ is not self-adjoint, one has $|F(\omega_{n}-\omega_{m})|=|F(\omega_{m}-\omega_{n})|$, but $|(H_1)_{n,m}| \neq |(H_1)_{m,n}|$ resulting in $W_{n,m} \neq W_{m,n}$. Another case is the one corresponding to a Hermitian perturbation operator $\hat{H}_1^\dag=\hat{H}_1$ but complex amplitude $f(t)$. In this case one has $|(H_1)_{n,m}|= |(H_1)_{m,n}|$  but $|F(\omega_{n}-\omega_{m})| \neq |F(\omega_{m}-\omega_{n})|$. An intriguing behavior is obtained when the frequency spectrum $F(\omega)$ of the perturbation is one sided. In particular, from Eq.(5) it readily follows that:\\
(i) If the spectrum $F(\omega)$  vanishes for positive (negative) frequencies, then $W_{n,m}=0$ for $\omega_m> \omega_n$ ($\omega_m < \omega_n$), while generally $W_{m,n} \neq 0$. Such a case corresponds to a perturbation-induced unidirectional transitions (maximal asymmetry).\\
(ii) If the spectrum $F(\omega)$  vanishes for frequencies $\omega>- \Omega$ (or for $\omega< \Omega$), with $\Omega > \omega_N-\omega_1$, then $W_{n,m}=0$ for any $m \neq n$. This means that the perturbation does not induce any transition and leaves the system in its original state.\par
As we show in the next section, the two above-mentioned properties persist in case of strong interaction, i.e. beyond the weak interaction limit $f(t) \rightarrow 0$. It should be mentioned that unidirectional transitions induced by a time-dependent perturbation with a one-sided Fourier spectrum represent a kind of temporal analogue of unidirectional wave scattering introduced by a spatial perturbation  with one-sided spatial Fourier spectrum disclosed in a few recent papers \cite{cazz,cazz0,cazz0bis,cazz1,cazz1bis,cazz2,cazz3,cazz4,cazz5,cazz6,cazz7}. In particular,  S.A.R. Horsley and collaborators showed on a rather general ground that a planar optical medium in which the real and imaginary parts of the dielectric permittivity are related one to another by spatial Kramers-Kronig relations is reflectionless for one incidence side \cite{cazz2}. In the spatial case, the scattering potential with one-sided Fourier spectrum, for example with vanishing negative spatial wave number components, cancels wave reflection from one incidence side because scattered waves cannot have wave numbers smaller than the one of the incidence plane wave. Similarly, in the temporal case a time-dependent perturbation with one-sided temporal Fourier spectrum can induce transition to e.g. higher energy levels, but not to lower energy levels \cite{cazz8}. Unlike the spatial analogue of Ref.\cite{cazz2}, where the scattering problem concerns the continuous spectrum of improper (non-normalizable) plane waves, in our case transitions occur between normalized (discrete) states of the Hamiltonian and the scattering problem can be formulated in a lower dimensional  space. An interesting application of unidirectional transitions in low-dimensional non-Hermitian systems is the explanation of the chiral behavior in the dynamical encircling of an EP, a phenomenon which is receiving a great attention in recent studies \cite{r21,r21bis,r23,r28,r35bis,r35}. This effect will be discussed in Sec.4.

\section{Unidirectional transitions and transitionless interactions}

\subsection{Unidirectional transitions} {\em Let $f(t)$ a complex function with one-sided Fourier spectrum, i.e. $F(\omega)=0$ for $\omega>0$, and let prepare the system at initial state $|n \rangle$, i.e. $c_{l}(-\infty)=\delta_{n,l}$. Then after the interaction, i.e. at $t \rightarrow \infty$, one has $c_m(\infty)=1$ for $m=n$, $c_m(\infty)=0$ (i.e. $W_{n,m}=0$) for any $m \neq n$ with $\omega_m \geq \omega_n$, and generally $c_m(\infty) \neq 0$ (i.e. $W_{n,m} \neq 0$) for $\omega_m < \omega_n$. In other words, a non-Hermitian interaction with negative Fourier spectrum does not induce transitions toward higher-energy states, regardless of the strength of the interaction and the form of $\hat{H}_0$ and $\hat{H}_1$.}\par
To prove the above property, beyond the weak perturbation limit $f \rightarrow 0$, let us introduce the amplitudes $a_l(t)=c_l(t) \exp [i \omega_n -\omega_l) t]$. From Eq.(2) it follows that $a_l$ satisfy the coupled equations
\begin{equation}
i \frac{d a_l}{dt}= (\omega_l-\omega_n) a_l + f(t) \sum_s (H_1)_{l,s} a_s.
\end{equation} 
Since the spectrum $F(\omega)$ vanishes for $\omega>0$, $f(t)= (1/ 2 \pi) \int_{-\infty}^{\infty} d \omega F(\omega) \exp(-i \omega t)$ is an analytic function of $t$ in the half complex plane ${\rm Im}(t) \geq 0$ with $|f(t)| \rightarrow 0$ as $|t| \rightarrow \infty$. Moreover, the real and imaginary parts of $f(t)$ are related each other by a Hilbert transform. Therefore, if we extend the variable $t$ into the complex plane, the solutions $a_l(t)$ to Eqs.(6) are holomorphic in the half complex plane ${\rm Im}(t) \geq 0$. Let us integrate Eq.(6) along the straight line $t= \xi +i \Delta$ of the analytic sector of the complex plane,  $\Delta>0$ and $-\infty < \xi < \infty$, with the initial condition $a_l( \xi \rightarrow -\infty)=\delta_{l,n}$. The solution $A_l( \xi, \Delta) \equiv a_l(t= \xi+i \Delta)$, with the initial condition $A_l(-\infty, \Delta)=\delta_{l,n}$, will depend parametrically on $\Delta$. Since $a_l(t)$ is holomorphic, one has 
\begin{equation}
\frac{\partial A_l}{\partial \Delta} =i \frac{\partial A_l}{\partial \xi}.
\end{equation}
 On the other hand, in the $\xi \rightarrow \infty$ limit and since $f(t=\xi+i \Delta) \rightarrow 0$ as $\xi \rightarrow \infty$,  from Eq.(6) the following asymptotic behavior of $A_l(\xi, \Delta)$ is found as $\xi \rightarrow \infty$
 \begin{equation}
 A_l(\xi, \Delta) \sim B_l (\Delta) \exp [-i (\omega_l-\omega_n) \xi]. 
 \end{equation}
  Combining Eqs.(7) and (8) yields
  \begin{equation}
  B_l(\Delta)=B_l(0) \exp[(\omega_l-\omega_n)\Delta].
  \end{equation}
  In particular, for $l=n$ one has $B_n(\Delta)=B_n(0)$, i.e. $B_n$ does not depend on $\Delta$.\\ 
   The transition probability $W_{n,m}=|c_m(\infty)|^2$, from state $|n \rangle$ to state $| m \rangle$, can be calculated as
   \begin{equation}
   W_{n,m}=|A_m( \xi= \infty, \Delta=0)|^2=|B_m(0)|^2=|B_m(\Delta)|^2 \exp[2 \Delta(\omega_n-\omega_m)]
   \end{equation}
 where we used Eqs.(8) and (9). In particular, we can compute $W_{n,m}$ by taking the limit $\Delta \rightarrow \infty$, i.e.
   \begin{equation}
   W_{n,m}= \lim_{\Delta \rightarrow \infty} |B_m(\Delta)|^2 \exp[2 \Delta(\omega_n-\omega_m)].
   \end{equation}
  In this limit, one has $f \rightarrow 0$ uniformly over the range $-\infty <  \xi < \infty$. Hence, from Eq.(6) one has $B_l(\Delta) \simeq \delta_{l,n}$ as $\Delta \rightarrow \infty$, since the interaction becomes vanishingly small, even thought it can be arbitrarily large at $\Delta=0$. Therefore, $c_n(\infty)=1$ whereas for any $m \neq n$ such that $\omega_m \geq \omega_n$ from Eq.(11) it follows that 
 $W_{n,m}=0$. Note that for $\omega_m<\omega_n$ the limit on the right hand side of Eq.(11) yields an indeterminate form ($0 \times \infty$), and thus rather generally $W_{n,m}$ can be nonvanishing.\\
 \\
 A similar property of asymmetric transitions holds when the spectrum $F(\omega)$ of the perturbation vanishes for negative (rather than positive) frequencies. In this case one has $W_{n,m}=0$ for any $m \neq n$ with $\omega_m \leq \omega_n$ while $W_{n,m}$ is generally nonvanishing for $\omega_m > \omega_n$. In other word, a non-Hermitian interaction with positive Fourier spectrum does not induce transitions toward lower-energy states, regardless of the strength of the perturbation.
 
 \subsection{Transitionless interactions}
  {\em Let $f(t)$ a complex function with vanishing Fourier spectrum $F(\omega)=0$ for $\omega>-\Omega$ (or for $\omega< \Omega$), with $\Omega \geq \omega_N-\omega_1$. Let us prepare the system in state $|n \rangle$ at $t= -\infty$, i.e. $c_{l}(-\infty)=\delta_{n,l}$. Then after the interaction, i.e. at $t \rightarrow \infty$, one has $c_m(\infty)=1$ for $m=n$ and $c_m(\infty)=0$ (i.e. $W_{n,m}=0$) for any $m \neq n$, regardless of the strength of the interaction and the form of $\hat{H}_0$ and $\hat{H}_1$ (transitionless interaction).}\par
To prove the above property beyond the weak perturbation limit $f \rightarrow 0$, let us assume, for the sake of definiteness, that the spectrum $F(\omega)$ of $f(t)$ vanishes for $\omega> -\Omega$, with $\Omega \geq \omega_N-\omega_1$. Note that, after setting $f(t)=g(t) \exp(i \Omega t)$, the spectrum $G(\omega)=F(\omega-\Omega)$ of $g(t)$ vanishes for positive frequencies $\omega>0$, and thus $g(t)$ is holomorphic in the half complex plane ${\rm Im}(t) \geq 0$, with $g(t) \rightarrow 0$ as $|t| \rightarrow \infty$. The evolution equations (2) of amplitudes $c_l(t)$ read
\begin{equation}
i \frac{dc_l}{dt}= g(t) \sum_s (H_1)_{l,s} c_s(t) \exp[i(\omega_l-\omega_s+\Omega)t]
\end{equation}
and $c_l(t)$ are analytic functions of $t$ in the the half complex plane ${\rm Im}(t) \geq 0$. Let us integrate Eq.(12) along the straight line $t=\xi+i \Delta$ of the complex plane, $\Delta>0$ and $-\infty < \xi < \infty$, with the initial condition $c_{l}(t=-\infty+i \Delta)= \delta_{n,l}$. The solution $c_l(t=\xi+i \Delta)$ is denoted by $C_l(\xi, \Delta)$ and depends parametrically on $\Delta$.  Owing to the analyticity of $c_l(t)$, one has
\begin{equation}
\frac{\partial C_l}{\partial \Delta} =i \frac{\partial C_l}{\partial \xi}.
\end{equation}
Since $g(t=\xi+i \Delta) \rightarrow 0$ as $\xi \rightarrow \infty$, from Eq.(12) it follows that $C_l(\xi, \Delta) \sim D_l(\Delta)$ as $ \xi \rightarrow \infty$, whereas  $C_l(\xi, \Delta) \sim \delta_{l,n}$ as $ \xi \rightarrow -\infty$. Note that Eq.(13) implies $dD_l/ d \Delta=0$, i.e. $D_l( \Delta)$ does not depend on $\Delta$. The transition probability $W_{n,m}$ can be thus calculated as
\begin{equation}
W_{n,m}=|C_m(\xi \rightarrow \infty, \Delta=0)|^2=|D_m(\Delta=0)|^2=|D_m(\Delta)|^2.
\end{equation}
Since $D_m(\Delta)$ does not depend on $\Delta$, we can take the limit $\Delta \rightarrow \infty$. In this limit, $g(t=\xi+i \Delta) \rightarrow 0$ and $|\exp[i(\omega_l-\omega_s+\Omega)(\xi+i \Delta) ]| \rightarrow 0$ uniformly in the range $-\infty <\xi < \infty$. The latter result follows from the fact that $\Omega$ is larger than any difference $(\omega_l-\omega_s)$ and because of $\Delta>0$. Therefore, for large $\Delta$, the amplitudes $C_l(\xi,\Delta)$ are decoupled and the solution to Eq.(12) is merely given by $C_l(\xi, \Delta) \simeq \delta_{l,s}$, i.e. $D_l(\Delta)=0$ for $m \neq n$ and $D_n(\Delta)=1$. Hence $W_{n,m}=0$ for $n \neq m$, i.e. the perturbation leaves the system unchanged as if the interaction had not occurred at all. Note that such a result holds regardless of the strength of the interaction and the precise form of $\hat{H}_0$ and $\hat{H}_1$, thus providing a nontrivial result  beyond the perturbative regime. We note that transitionless interactions, beyond the perturbative regime, can be found also in certain Hermitian models, for example in the optical Bloch equations for a driven two-level atom describing the transition between the
two atomic levels induced by a nearly resonant optical pulse , where for special areas of the pulse the optical field leaves the atom in its original state \cite{All1,All2}. However, in the non-Hermitian case considered in the present work the transitionless effect is a much more general phenomenon that occurs regardless of the number of levels, amplitude of the perturbation and specific form of $\hat{H}_0$ and $\hat{H}_1$.  
 
 \begin{figure}[b]
\includegraphics[width=12cm]{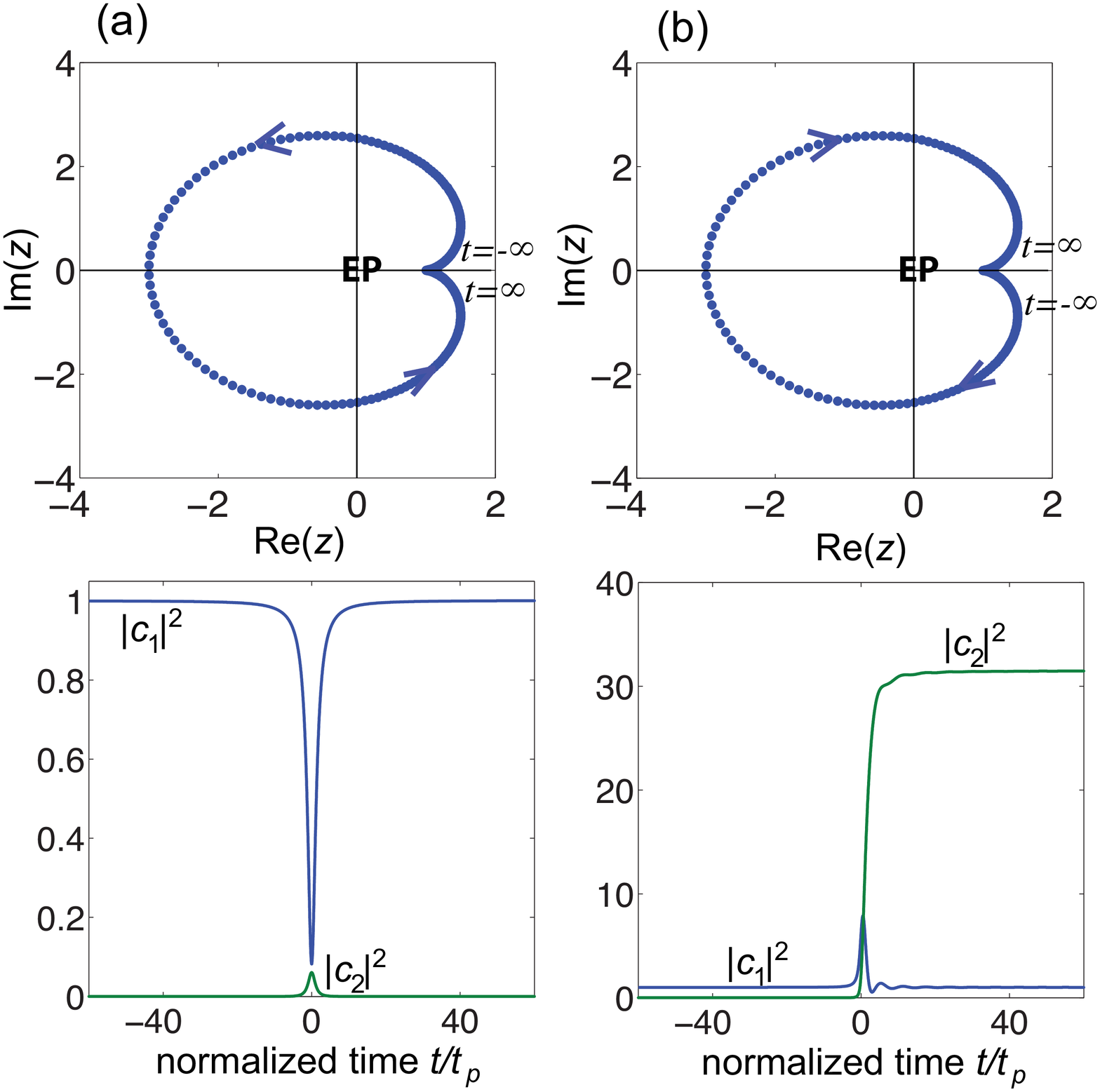}
\caption{(Color online). Dynamical encircling of an EP for a two-level system. (a) Counter-clockwise loop described by $z(t)=1+f(t)$, with $f(t)$ defined by Eq.(23) with $A=1$ and $t_p=-0.5$. The numerically-computed evolution of the probabilities $|c_1(t)|^2$ and $|c_2(t)|^2$ is shown in the lower panel. The system is prepared in state $|1 \rangle$ at $t= -\infty$. (b) Same as (a) but for a clockwise circulation of the loop ($A=1$, $t_p=0.5$). Note that, while in (a) the system remains in its initial state $|1 \rangle$, in (b) a flip to state $|2 \rangle$ is observed, corresponding to a chiral behavior of EP encircling.} 
\end{figure}

 \begin{figure}[b]
\includegraphics[width=12cm]{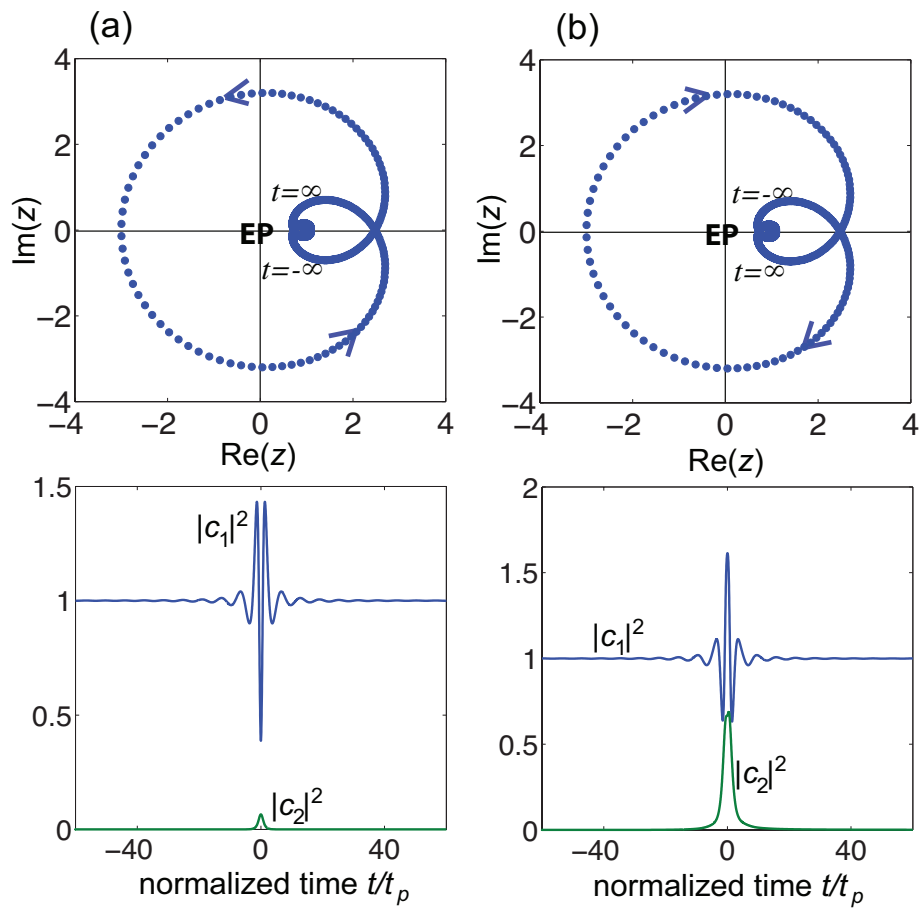}
\caption{(Color online). Same as Fig.1, bur for $f(t)$ defined by Eq.(24) with $\Omega \tau_p>0$. In (a) $A=1$ and $t_p=-0.5$ and $\Omega=-2$, corresponding to counter-clockwise circulation of the loop. In (b) $A=1$ and $t_p=0.5$ and $\Omega=2$, corresponding to clockwise circulation of the loop. Note that in this case the chiral behavior disappears since the system remains in its initial state $|1 \rangle$ regardless of the circulation direction of the loop.} 
\end{figure}

 \begin{figure}[b]
\includegraphics[width=12cm]{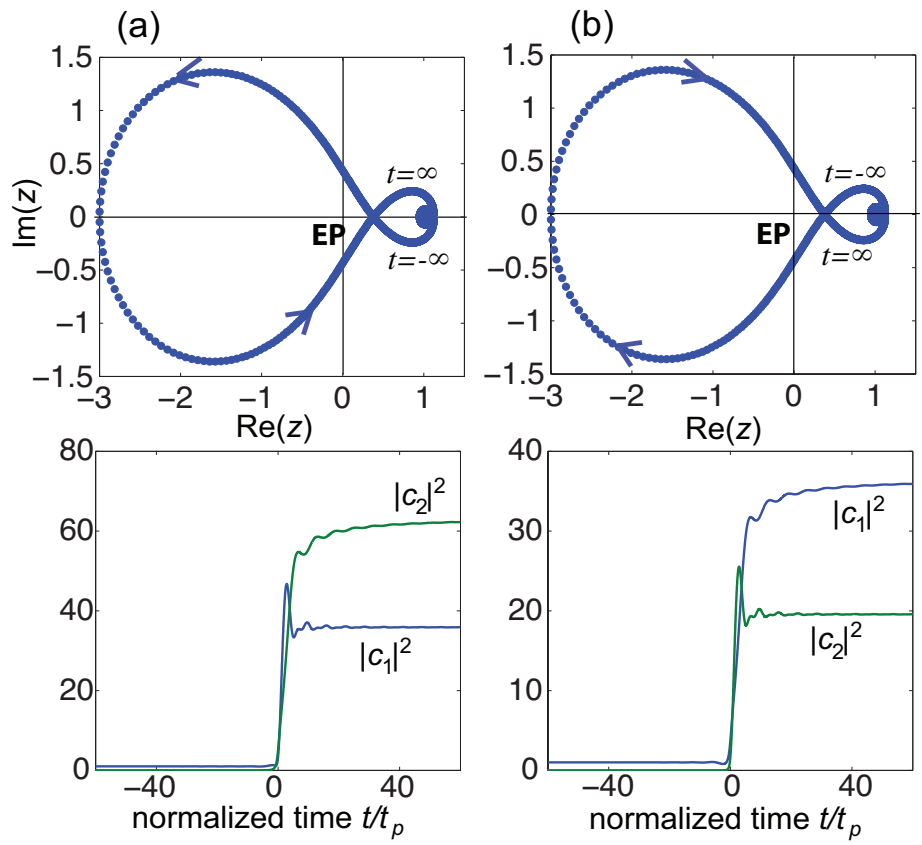}
\caption{(Color online). Same as Fig.2, bur for reversed sign of $\Omega \tau_p$ [$t_p=-0.5$, $\Omega=2$ in (a), $t_p=0.5$, $\Omega=-2$ in (b)]. } 
\end{figure}
 
 \begin{figure}[b]
\includegraphics[width=12cm]{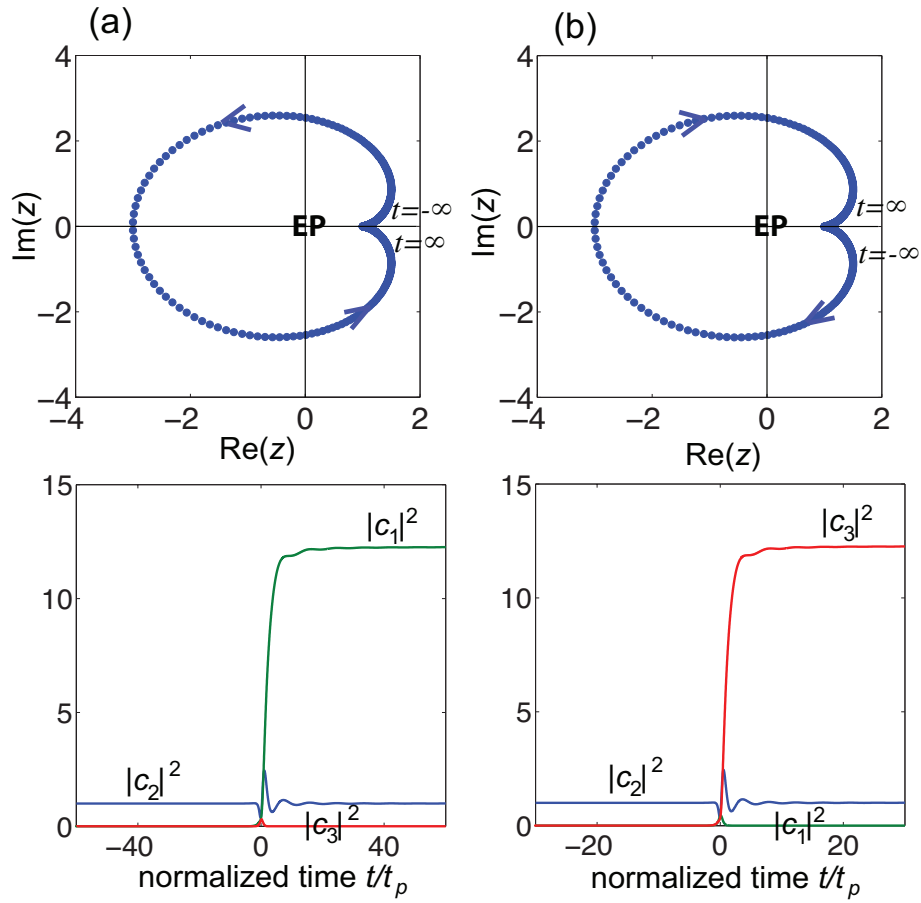}
\caption{(Color online). Dynamical encircling of a third-order EP in a three-level system. (a) Counter-clockwise loop described by $z(t)=1+f(t)$, with $f(t)$ defined by Eq.(23) with $A=1$ and $t_p=-0.5$. The numerically-computed evolution of the probabilities $|c_1(t)|^2$, $|c_2(t)|^2$ and $|c_3(t)|^2$ is shown in the lower panel. The system is prepared in state $|2 \rangle$ at $t= -\infty$. (b) Same as (a) but for a clockwise circulation of the loop ($A=1$, $t_p=0.5$). Note that, while in (a) the system moves to state $| 1 \rangle$, in (b) the system moves toward state $|3 \rangle$, indicating a chiral behavior.} 
\end{figure}

 \begin{figure}[b]
\includegraphics[width=12cm]{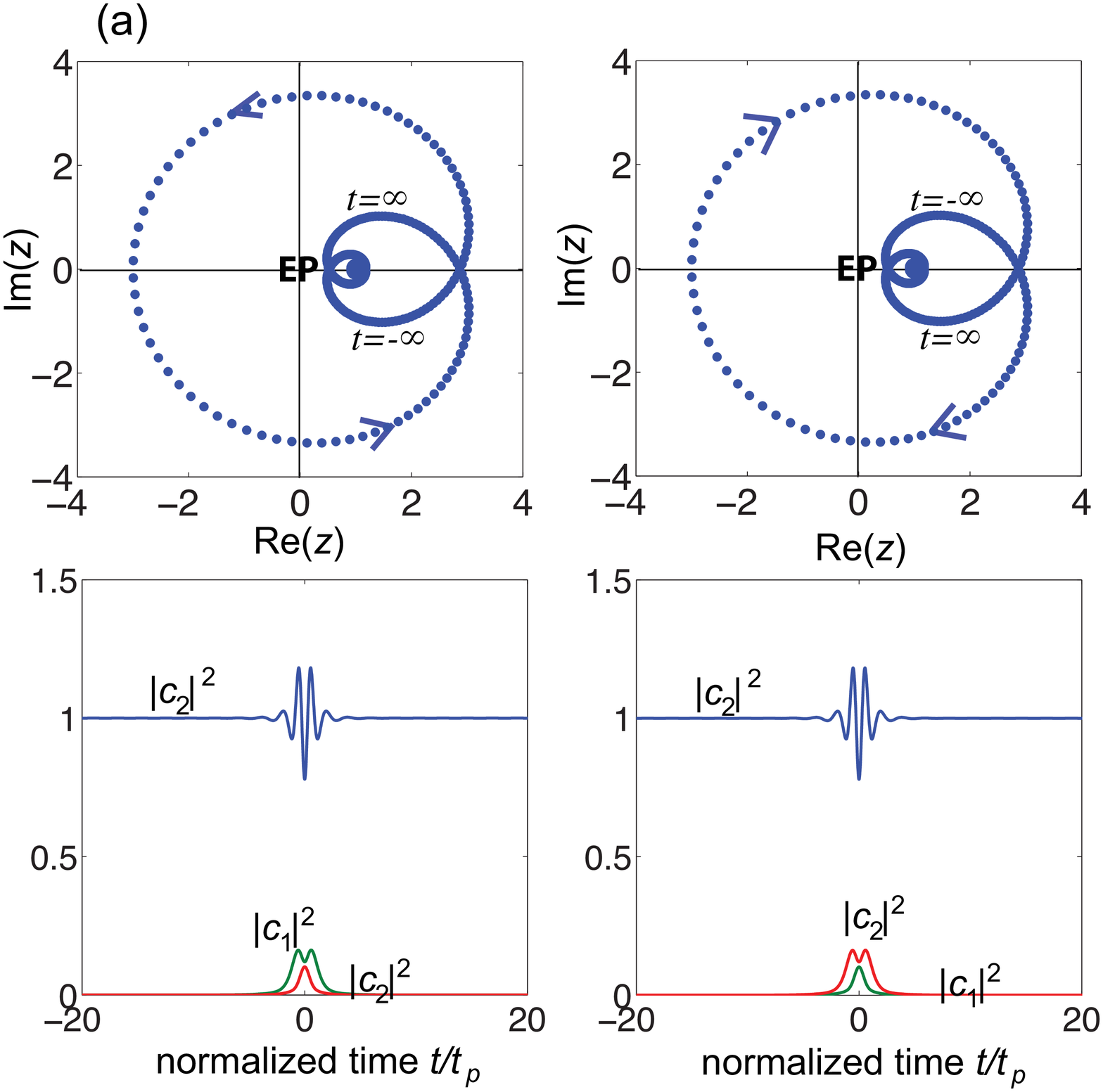}
\caption{(Color online). Same as Fig.4, bur for $f(t)$ defined by Eq.(24). In (a) $A=1$, $t_p=-0.5$ and $\Omega=-2 \sqrt{2}$, corresponding to counter-clockwise circulation of the loop. In (b) $A=1$ and $t_p=0.5$ and $\Omega=2 \sqrt{2}$, corresponding to clockwise circulation of the loop. Note that in this case the chiral behavior disappears since the system remains in its initial state $|1 \rangle$ regardless of the circulation direction of the loop. A similar behavior is observed by lowering $| \Omega |$ down to $\sqrt{2}$.} 
\end{figure}

\section{An application: dynamical encircling of an exceptional point}
One of the most interesting scenario of a non-Hermitian Hamiltonian is the coalescence of two (or more) eigenvalues and the corresponding eigenvectors at so-called
exceptional points \cite{r31,r32,r33}, as opposed to the diabolic
point  degeneracy of Hermitian operators, at which
the eigenvalues coalesce while the eigenvectors remain different. EPs have attracted considerable attention in recent years as a peculiar signature of non-Hermitian systems, especially in connection with the dynamical properties associated to the encircling of an EP when a parameter of the system is periodically varied. These include state-flip or the accumulation of a geometric phase for very slow circling \cite{r15,r34,r36,r37} and chiral behavior associated to breakdown of adiabaticity for faster circling \cite{r21,r23,r28,r35}. In particular, non-adiabatic transitions leading to a chiral behavior, i.e. selection of a different final state when encircling an
EP in a clockwise or a counter-clockwise direction, have been recently demonstrated in an experiment using engineered smoothly deformed metallic waveguides \cite{r35}. Here we discuss dynamical encircling properties of an EP in the framework of non-Hermitian time-dependent perturbation theory presented in the previous sections and show how different final state selection results from asymmetric transition probabilities or from transitionless encircling. The analysis is exemplified by considering EP of second- (EP2) and third-order (EP3) in two and three-level systems, however the method can be extended to EPs of higher order \cite{Gra,PRX}.

\subsection{Encircling an exceptional point: two-level system}
Let us consider a two-level system which is described, in the two-level state basis $|I \rangle$ and $II \rangle$, by the time-dependent Hamiltonian (1) with
\begin{equation}
\hat{H}_0= | I \rangle \langle II |+ | II \rangle \langle I| \;, \;\; \hat{H}_1= | II \rangle \langle I |,
\end{equation}
i.e.
\begin{equation}
\hat{H}(t)=| I \rangle \langle II |+ z(t) | II \rangle \langle  I |
\end{equation}
where we have set
\begin{equation}
z(t) \equiv 1+f(t)
\end{equation}
and $z(t) \rightarrow 1 $ as $t \rightarrow \pm \infty$. Note that, after setting $ | \psi(t) \rangle= \psi_1(t) | I \rangle + \psi_2(t) | II \rangle$, the Schr\"odinger equation for the amplitudes $\psi_1$ and $\psi_2$ in the two-level basis reads
\begin{eqnarray}
i \frac{d \psi_1}{dt} & = & \psi_2 \\
i \frac{d \psi_2}{dt} & = & z(t) \psi_1. 
\end{eqnarray}
The properties of Eqs.(18) and (19) when $z(t)$ describes a closed loop in complex plane have been investigated in great details in Ref.\cite{r21} for some exactly integrable cases. Note that the origin $z=0$ is an EP of second order (EP2), since the $2 \times 2$ matrix associated to $\hat{H}$ is a Jordan normal form at $z=0$. The instantaneous eigenvalues of $\hat{H}$ are given by $\omega_{1,2}(t)= \pm \sqrt{z(t)}$. The branch-point at $z=0$ implies that, when $z(t)$ describes a closed loop around $z=0$, starting from $z=1$ at $t \rightarrow -\infty$ and ending at the same point $z=1$ at $t \rightarrow \infty$, the two eigenvalues have exchanged, and also the eigenvectors up to a constant factor (state flip) \cite{r21}. The state flip and eigenvalue exchange is not found when the loop does not encircle the EP. State flip in the former case requires adiabatic following. However, recent works have shown that non-Hermitian dynamics can easily break the adiabatic limit \cite{r21,r23,r28,r35} owing to Stokes phenomenon of asymptotics \cite{r21}. In particular, dynamical encircling of an EP can show a chiral behavior \cite{r21,r23,r35}, i.e. the final state depends on the circulation direction of the loop. Here we consider dynamical EP encircling described by a complex function $z(t)$ given by Eq.(17) with $f(t)$ satisfying the analytic conditions discussed in the previous section, and show how the chiral behavior can be readily explained in terms of asymmetric transition probabilities $W_{1,2} \neq W_{2,1}$. The eigenstates $|1 \rangle$ and $|2 \rangle$ of the unperturbed Hamiltonian $\hat{H}_0$, with eigenvalues $\omega_1=-1$ and $\omega_2=1$, are given by 
\begin{equation}
|1 \rangle= \frac{1}{\sqrt{2}} \left( | I \rangle - | II \rangle \right) \; , \; \;  |2 \rangle= \frac{1}{\sqrt{2}} \left( | I \rangle + | II \rangle \right).
\end{equation} 
In the $ \{ |1 \rangle, |2 \rangle \}$ basis, i.e. after setting $| \psi(t)= c_1 (t) \exp(-i \omega_1 t) | 1 \rangle+ c_2 (t) \exp(-i \omega_2 t) | 2 \rangle$, the Schr\"odinger equation for the amplitudes $c_1$ and $c_2$ [Eq.(2)] reads explicitly
\begin{eqnarray}
i \frac{dc_1}{dt} & = & -\frac{z(t)-1}{2}c_1-\frac{z(t)-1}{2} c_2 \exp(-2it) \\
i \frac{dc_2}{dt} & = & \frac{z(t)-1}{2}c_1 \exp(2it)+\frac{z(t)-1}{2} c_2  
\end{eqnarray}
At initial time $t \rightarrow -\infty$ let us prepare the system in one of its eigenstates, for example in eigenstate $|1 \rangle$, i.e. let us assume $c_1(-\infty)=1$ and $c_2(-\infty)=0$. The perturbation function $f(t)$ is chosen so as $z(t)=1+f(t)$ describes a closed loop in complex plane encircling once the EP at $z=0$. Let us consider, as a first example, the case where $f(t)$ is of the form
\begin{equation}
f(t)=\frac{A}{(t-i t_p)^2}
\end{equation}
which is a meromorphic function with a pole on the imaginary axis at $t=i t_p$ ($t_p$ real). Parameter values $A$ and $t_p$ are chosen so as a single loop, circling around $z=0$,  is obtained when time $t$ varies from $t=-\infty$ to $t= \infty$. Note that by changing the sign of $t_p$, i.e.mirror-reversing the position of the pole with respect to the real axis, the circulation direction of the loop is reversed; see Fig.1. For $t_p<0$ the loop is traversed counterclockwise [Fig.1(a)]; in this case $f(t)$ is holomorphic in the ${\rm Im}(t) \geq 0$ half complex plane and its spectrum $F(\omega)$ vanishes for $\omega>0$. According to the result of Sec.3.1, the perturbation is not able to induce any transition and one has $c_1(\infty)=1$ and $W_{1,2}=0$, i.e. after the cycle the system has remained in its initial state. On the other hand, for $t_p>0$ the loop is traversed clockwise [Fig.1(b)], $f(t)$ is holomorphic in the ${\rm Im}(t) \leq 0$ half complex plane and its spectrum $F(\omega)$ vanishes for $\omega<0$. In this case one has $c_1(\infty)=1$ but the perturbation can induce a transition, i.e. $W_{12} \neq 0$; in particular for parameter values used in Fig.1(b) one has $|c_2(\infty)|^2 \simeq 31.47$, which is much larger than one: this means that a state flip has occurred by traversing the loop, from state $|1 \rangle$ to (almost) state $|2 \rangle$. Therefore, the chiral behavior observed when encircling the EP clockwise or counterclockwise stems from the asymmetric transition probability induced by the one-sided spectrum perturbation. As a second example, let us consider the case where $f(t)$ has the form
\begin{equation}
f(t)=\frac{A }{(t-i t_p)^2} \exp(-i \Omega t)
\end{equation}
which differs from Eq.(23) for the additional exponential term $\exp(-i \Omega t)$. Again we chose parameters $A$, $t_p$ and $\Omega$ so as the EP at $z=0$ is encircled once when $t$ varies from $t= - \infty$ to $t= \infty$. Note that, by reversing the sign of both $t_p$ and $\Omega$, the circulation direction of the loop is reversed; see Fig.2. For $\Omega \tau_p>0$ and  $|\Omega| \geq \omega_2-\omega_1=2$, the theorem of Sec.3.2 is satisfied so that, regardless of the circulation direction of the loop, no transitions should occur ($W_{1,2}=0$ for either $\Omega>2$ and $\Omega<-2$). Indeed, numerical results shown in Fig.2 indicate that in this case the chiral behavior previously observed vanishes and encircling the EP clockwise or counterclockwise does not induce any state transition. Interestingly, for $\Omega \tau_p<0$ the conditions of the theorem stated in Sec.3.2 are not met, and encircling the EP yields a different dynamical scenario since transitions are now allowed. An example of numerical results is shown in Fig.3. Note that in this case asymmetric dynamics for clockwise and counter-clockwise circulation direction of the loop is observed, however as compared to the case of Fig.1 adiabatic following is broken for both circulation directions since a mixtrure of the two adiabatic states is obtained after one encircling of the EP.

\subsection{Encircling an exceptional point: three-level system}
In a three-level system EP of third order (EP3) can be found \cite{Hei,Graefe, Kim,Heiss,Wunner}. Let us consider, as an example, the three-level system described, in the three-level state basis $ \{ |I \rangle, II \rangle, | III \rangle \}$, by the time-dependent Hamiltonian (1) with
\begin{equation}
\hat{H}_0= | I \rangle \langle II |+ | II \rangle \langle I| + | II \rangle \langle III | + | III \rangle \langle II | \; , \;\;
\hat{H}_1= | II \rangle \langle I | + | III \rangle \langle II|
\end{equation}
i.e.
\begin{equation}
\hat{H}(t)=| I \rangle \langle II |+ z(t) | II \rangle \langle I| + | II \rangle \langle III | + z(t) | III \rangle \langle II | 
\end{equation}
where we have set
\begin{equation}
z(t) \equiv 1+f(t)
\end{equation}
and $z(t) \rightarrow 1 $ as $t \rightarrow \pm \infty$. After setting $ | \psi(t) \rangle= \psi_1(t) | I \rangle + \psi_2(t) | II \rangle+\psi_3(t) | III \rangle$, the Schr\"odinger equation for the amplitudes $\psi_1$, $\psi_2$ and $\psi_3$ in the three-level basis reads
\begin{eqnarray}
i \frac{d \psi_1}{dt} & = & \psi_2 \\
i \frac{d \psi_2}{dt} & = & z(t) \psi_1+ \psi_3 \\
i \frac{d \psi_3}{dt} & = & z(t) \psi_2 . 
\end{eqnarray}
 Note that the origin $z=0$ is an EP of third order (EP3), since the $3 \times 3$ matrix associated to $\hat{H}$ is a Jordan normal form at $z=0$. A rather general theory of the cyclic quasi-static evolution of eigenavlues and eigenvectors for a third-order EP has been presented in Refs. \cite{Hei,Graefe}. In our example, the instantaneous eigenvalues of $\hat{H}$ are given by $\omega_{1}(t)= - \sqrt{2 z(t)}$, $\omega_2(t)=0$ and $\omega_3(t)= \sqrt{2 z(t)}$, i.e. one eigenvalue is constant (like in the case of bottom Fig.1 in Ref.\cite{Graefe}). The branch-point at $z=0$ implies that, when $z(t)$ describes a closed loop around $z=0$, starting from $z=1$ at $t \rightarrow -\infty$ and ending at the same point $z=1$ at $t \rightarrow \infty$, the two eigenvalues $\omega_1$ and $\omega_3$ have exchanged, and also the eigenvectors up to a constant factor (state flip). On the other hand, the eigenvalue $\omega_2(t)=0$ and corresponding eigenstate is not changed after the cycle. Like in the two-level case, such a dynamical scenario can be broken owing to non-adiabatic effects, and a chiral behavior can be observed. To highlight the chiral behavior of EP3 in the framework of the time-dependent perturbation theory developed in the previous section, let us write the Hamiltonian $\hat{H}(t)$ on the basis $|n \rangle$ ($n=1,2,3$) of the unperturbed and Hermitian Hamiltonian $\hat{H}_0$. The eigenstates $|1 \rangle$, $|2 \rangle$ and $|3 \rangle$ of $\hat{H}_0$, with eigenvalues $\omega_1=- \sqrt{2}$, $\omega_2=0$ and $\omega_3= \sqrt{2}$, are given by 
\begin{equation}
|1 \rangle= \frac{1}{2} \left( | I \rangle -  \sqrt{2} | II \rangle + | III \rangle \right) \; , \;   |2 \rangle= \frac{1}{\sqrt{2}} \left( | I \rangle - | III \rangle \right) \; , \; 
|3 \rangle= \frac{1}{2} \left( | I \rangle +  \sqrt{2} | II \rangle + | III \rangle \right).
\end{equation} 
In the $ \{ |1 \rangle, |2 \rangle , | 3 \rangle \}$ basis, i.e. after setting $| \psi(t)= c_1 (t) \exp(-i \omega_1 t) | 1 \rangle+ c_2 (t) \exp(-i \omega_2 t) | 2 \rangle+c_3(t) \exp(-i \omega_3 t) | 3 \rangle$, the Schr\"odinger equation for the amplitudes $c_1$, $c_2$ and $c_3$ [Eq.(2)] reads explicitly
\begin{eqnarray}
i \frac{dc_1}{dt} & = & -\frac{z(t)-1}{\sqrt{2}}c_1-\frac{z(t)-1}{2} c_2 \exp(-i \sqrt{2} t) \\
i \frac{dc_2}{dt} & = & \frac{z(t)-1}{2}c_1 \exp(i \sqrt{2}t)- \frac{z(t)-1}{2}c_3 \exp(-i \sqrt{2}t) \\
i \frac{dc_3}{dt} & = & \frac{z(t)-1}{\sqrt{2}}c_3+\frac{z(t)-1}{2} c_2 \exp(i \sqrt{2} t) 
\end{eqnarray}
To show breakdown of adiabatic theorem and chirality when encircling the third-order EP, let us prepare the system at initial time $t \rightarrow -\infty$  in the eigenstate $|2 \rangle$, i.e. let us assume $c_2(-\infty)=1$ and $c_1(-\infty)=c_3(-\infty)=0$. The perturbation function $f(t)$ is chosen so as $z(t)=1+f(t)$ describes a closed loop in complex plane encircling the EP at $z=0$. Let us consider, as a first example, the case where $f(t)$ is of the form (23). Like in the two-level problem discussed above, parameter values $A$ and $t_p$ are chosen so as a single loop, circling around $z=0$,  is obtained when time $t$ varies from $t=-\infty$ to $t= \infty$. Note that by changing the sign of $t_p$, i.e. reversing the position of the pole, the circulation direction of the loop is reversed; see Fig.4. For $t_p<0$ the loop is traversed counter-clockwise [Fig.4(a)]; in this case $f(t)$ is holomorphic in the ${\rm Im}(t) \geq 0$ half complex plane and its spectrum $F(\omega)$ vanishes for $\omega>0$. According to the theorem of Sec.3.1, the perturbation can induce a transition to the lower-energy state $|1 \rangle$, but not to the upper-energy level $|3 \rangle$, i.e. $W_{2,3}=0$, $W_{2,1} \neq 0$ and $c_2(\infty)=1$. For parameter values used in the simulations of Fig.4(a), one has $|c_{1}(\infty)|^2 \simeq 12.26$, which is much larger than one: this means that after the cycle, traversed in the counter-clockwise direction, the system flips into the lower-energy state $|1 \rangle$. This state flip shows that non adiabatic effects arise in dynamical encircling of the EP.  On the other hand, for $t_p>0$ the same loop is traversed clockwise [Fig.4(b)], $f(t)$ is holomorphic in the ${\rm Im}(t) \leq 0$ half complex plane and its spectrum $F(\omega)$ vanishes for $\omega<0$. In this case according to the theorem of Sec.3.1 one has $c_2(\infty)=1$, $W_{2,1}=0$  and $W_{2,3} \neq 0$, i.e. the perturbation can induce this time a transition to the upper-energy level $| 3 \rangle$: as compared to the case of Fig.4(a), the role of levels $|1 \rangle$ and $|3 \rangle$ is exchanged, i.e. a flip to state $|3 \rangle$ has occurred by traversing the loop in the clockwise direction. Therefore, a chiral behavior is observed when encircling the EP clockwise or counterclockwise, which results from the asymmetric transition probability rates. Note that the chiral behavior observed in this case is different than the one found for the two-level model discussed in the previous section (Fig.1): in fact, in the two-level system the chirality arises because of breakdown of the adiabatic following when the loop is traversed in one direction, but not in the opposite one. In the three-level system adiabatic following is broken when the loop is traversed in both directions, and the final state always differs than the initial one.\\
 As a second example, let us consider the case where $f(t)$ has the form (24).  Again we chose parameters $A$, $t_p$ and $\Omega$ so as the EP at $z=0$ is encircled once when $t$ varies from $t= - \infty$ to $t= \infty$. Note that, by reversing the sign of both $t_p$ and $\Omega$, the circulation direction of the loop is reversed; see Fig.5. For $\Omega \tau_p>0$ and  $|\Omega| \geq \omega_3-\omega_1=2 \sqrt{2}$, the theorem of Sec.3.2 holds so that, regardless of the circulation direction of the loop, no transitions should occur ($W_{2,3}=W_{2,1}=0$ for both $\Omega>2 \sqrt{2}$ and $\Omega<-2 \sqrt{2}$  \footnote{Note that, since the initial state is the middle energy state $|2 \rangle$,  no transition is found for a smaller value of $| \Omega|$, but larger than $\sqrt{2}$.}). Indeed, numerical results shown in Fig.5 indicate that in this case the chiral behavior previously observed vanishes and encircling the EP clockwise or counterclockwise does not induce any state transition.

\section{Conclusions}

In this work we have extended the ordinary time-dependent perturbation theory of quantum mechanics to the non-Hermitian realm by considering transitions in a stationary Hermitian system induced by a non-Hermitian perturbation. While the ordinary (Hermitian) theory predicts that the transition probabilities induced by a weak perturbation are symmetric with respect to exchange of initial an final states, for a non-Hermitian perturbation the transition probabilities generally turn out to be asymmetric when initial and final states are reversed. In particular, for a time-dependent perturbation with one-sided Fourier spectrum, i.e. with only positive (or negative) frequency components, transitions are fully unidirectional. By use of complex analysis and properties of holomorphic functions,  we showed that such a highly-asymmetric behavior is an exact result that holds even for a strong interaction, i.e. beyond the perturbative regime. Interestingly, strong non-Hermitian interactions can be tailored to be  transitionless, i.e. the perturbation leaves the system unchanged as if the interaction had not occurred at all. Such a result is a rather general one, independent of the specific form of $\hat{H}_0$ and $\hat{H}_1$, and thus very distinct than transitionless interactions found in special Hermitian models \cite{All1,All2}. As an application of the general theory, we discussed breakdown of adiabatic theorem and chirality of exceptional point encircling, showing how the chiral behavior, i.e. different final state depending on the circulation direction of the loop, is the signature of asymmetric transition probabilities. The present results shed new light into the dynamical behavior of non-Hermitian systems, revealing how non-Hermitian perturbations can be tailored to induce selective transitions in a stationary system.

\end{document}